\documentstyle[aps,epsf,prl,epsfig,twocolumn]{revtex}

\begin{document}

\title{Quantum fluctuations of a vortex in 
         a dilute Bose-Einstein condensate }

\author{ Jacek Dziarmaga and Jakub Meisner }

\address{
Instytut Fizyki Uniwersytetu Jagiello\'nskiego, 
ul.~Reymonta 4, 30-059 Krak\'ow, Poland 
}

\date{ October 26, 2004 }

\maketitle

\begin{abstract}
A vortex in a quasi two-dimensional Bose-Einstein condensate is subject
to the Magnus force and can be effectively described as a planar
particle in a uniform magnetic field. Quantization of this
effective particle leads to the lowest Landau level where the most
localized wave function is a gaussian. In this gaussian state
vortex position seems to fluctuate with an average magnitude set
by the magnetic width of the gaussian. We readdress this problem
using the number-conserving version of the Bogoliubov theory. We
find that the Bogoliubov mode that might be interpreted as a
fluctuation of vortex position actually does not contribute to 
position fluctuation at all. The only non-zero contribution comes 
from phonons but it is an order of magnitude less than the simple
estimate, based on the magnetic width of the effective gaussian
wave packet. 
\end{abstract}

%%%%%%%%%%%%%%%%%%%%%%%%%%%%%%%%%%%%%%%%%%%%%%%%%%%%%%%%%%%%%%%%%%%%%%%%%%%%%%%%
%%%%%%%%%%%%%%%%%%%%%%%%%%%%%%%%%%%%%%%%%%%%%%%%%%%%%%%%%%%%%%%%%%%%%%%%%%%%%%%%
\section{ Introduction }
%%%%%%%%%%%%%%%%%%%%%%%%%%%%%%%%%%%%%%%%%%%%%%%%%%%%%%%%%%%%%%%%%%%%%%%%%%%%%%%%

Bose-Einstein condensation of dilute alkali gases \cite{nobel} opens up
new possibilities to investigate Bose-Einstein-condensed systems under
flexible and well controlled conditions. For example it is possible to
manipulate with the dimensionality of the trap to achieve quasi
one or two-dimensional condensates \cite{12D}. Temperatures
can be reduced to the pK range \cite{pK} bringing the system close to
the quantum regime. Various manifestations of superfluidity have been
subject of intensive research, the most spectacular have been probably
vortices \cite{vortex} and Abrikosov vortex lattices \cite{lattice} in
rotating traps. It is predicted \cite{QHE} that further increase in the
rotation speed of a quasi-two-dimensional trap can drive the system from
the regime of BEC to the regime of strongly correlated bosonic quantum
Hall liquids with even denominator filling fractions.

A dilute bosonic gas in a rotating two-dimensional trap at very low
temperature is a setup required for quantum Hall liquids but at the same
time it can be used to experimentally test quantum fluctuations of a 2D
vortex in a dilute Bose-Einstein condensate. In fact melting of the
Abrikosov vortex lattice driven by quantum fluctuations of vortices is the first step
from a Bose-Einstein condensate towards strongly correlated quantum Hall
liquids. In this paper we develop a theory of quantum vortex fluctuations.
  
%%%%%%%%%%%%%%%%%%%%%%%%%%%%%%%%%%%%%%%%%%%%%%%%%%%%%%%%%%%%%%%%%%%%%%%%%%%%%%%%
\section{ Vortex fluctuations }
%%%%%%%%%%%%%%%%%%%%%%%%%%%%%%%%%%%%%%%%%%%%%%%%%%%%%%%%%%%%%%%%%%%%%%%%%%%%%%%%

An effective action for a planar vortex has been worked out in
Ref.\cite{Thouless}. In terms of vortex position on the plane $\vec
X=({\cal X},{\cal Y})$ the action reads
\begin{equation}
{\cal S}_{\rm eff}~=~
\int dt~
\left[
\frac{m_{\rm eff}}{2} \dot{\vec{X}}^2 +
\frac{\kappa}{2} \hbar\rho~ 
\hat{e}_z \left(\vec{X}\times\dot{\vec{X}}\right)
\right]~.
\label{Seff}
\end{equation}
Here $m_{\rm eff}$ is an effective mass of the vortex that depends
on the size of the system, $\rho$ is planar density of atoms in a
uniform condensate, the integer $\kappa$ is the winding number
of the vortex, and $\hat{e}_z$ is a unit vector along the $z$-axis
normal to the plane. Formally this is the action of a planar particle 
in a uniform magnetic field normal to the plane, as we can see from 
its Euler-Lagrange equation:  
$m_{\rm eff}\ddot{\vec{X}}=\kappa\hbar\rho~\dot{\vec{X}}\times\hat{e}_z$.  
The effective Lorenz force on the right hand side is the Magnus
force. Because vortices with $|\kappa|>1$ are unstable, in the
following we consider only $\kappa=+1$.

The effective particle can be quantized and its eigenstates can be
classified in the degenerate Landau levels. The energy gap between
Landau levels is $\frac{\hbar\rho}{m_{\rm eff}}$ so in the limit
of vanishing effective mass the spectrum can be truncated to the
lowest Landau level (LLL). The degenerate eigenmodes of the LLL
can be chosen as $({\cal Z}^{*})^l~e^{-\frac{1}{2}\rho{\cal Z}^*{\cal
Z}}~$. Here ${\cal Z}={\cal X}+i\;{\cal Y}$ is a complex position 
of the vortex. We get the most localized eigenstate when the non-negative 
integer $l=0$:
\begin{equation}
e^{ - \frac12 \rho {\cal Z}^*{\cal Z}}=e^{- \frac12 \rho ({\cal X}^2+{\cal Y}^2)}~.
\label{gauss_eff}
\end{equation}
This gaussian seems to determine the quantum fluctuation
of the vortex position.

In the next Section we rederive this result in the framework of
the particle-number-conserving Bogoliubov theory \cite{NBT}, but 
then we use the same Bogoliubov theory to show that this result 
is incomplete and even misleading.
 
%%%%%%%%%%%%%%%%%%%%%%%%%%%%%%%%%%%%%%%%%%%%%%%%%%%%%%%%%%%%%%%%%%%%%%%%%%%%%%%%
\section{ Vortex fluctuations       \\ 
          in the Bogoliubov theory   }
%%%%%%%%%%%%%%%%%%%%%%%%%%%%%%%%%%%%%%%%%%%%%%%%%%%%%%%%%%%%%%%%%%%%%%%%%%%%%%%%

In the mean-field approximation a stationary condensate wave
function is a solution of the time-independent Gross-Pitaevskii
equation 
\begin{equation}
\mu\phi=
-\frac{\hbar^2}{2m}\nabla^2\phi+
Ng_0 |\phi|^2\phi+V\phi~.
\end{equation} 
Here $g_0$ is an effective 2D interaction strength and $N$ is a total, conserved number of atoms in the box. 
In this section we ignore the trap potential $V$ and consider a uniform 2D
condensate in a box. In the strong interaction Thomas-Fermi (TF) regime ($Ng_0\equiv g\gg 1)$ the uniform
condensate has a wave function $\phi=\sqrt{\rho/N}$ with $\mu=\rho g_0$ . 
In this 
case vortex has a wave function
$\phi_0(r,\theta)=f(r)\exp(i\theta)$ in polar coordinates $r$ and
$\theta$, where the modulus $f(r)$ interpolates between $f(0)=0$ and
$f(\infty)=\sqrt{\rho/N}$.

Small fluctuations around the vortex can be expanded into
Bogoliubov modes with complex amplitudes $b_m$
\begin{equation}
\phi_b(\vec x)=
{\mathcal N}
\left[
\phi_0(\vec x)+
\sqrt{\frac{2}{N}}
\sum_m 
b_m u_m(\vec x)+b_m^* v^*_m(\vec x)
\right]~,
\label{expansion}
\end{equation}
where 
\begin{eqnarray}
{\mathcal N}^{-1}(b,b^*)&=&
\nonumber\\
1+\frac{2}{N} &\sum_{m,n}&
b_m^*b_n\langle u_m|u_n\rangle+
b_mb_n \langle v_m^*|u_n\rangle+
\nonumber\\
&& 
b_m^*b_n^*\langle u_m|v_n^*\rangle+
b_mb_n^*\langle v_n|v_m\rangle
\end{eqnarray} 
is the normalization factor, such that $\left<\phi_b|\phi_b\right>=1$.
In addition to the usual phonon modes with the characteristic
phonon dispersion relation, there is a normalizable mode with zero 
frequency 
\begin{eqnarray}
u_0
&=&
-\sqrt{\frac{N}{\rho}}~
\frac{\partial}{\partial z}\phi_0
~=~
-\sqrt{\frac{N}{\rho}}~
\left(
\frac{df}{dr}+\frac{f}{r}
\right)
~,
\nonumber\\
v^*_0&=&
-\sqrt{\frac{N}{\rho}}~
\frac{\partial}{\partial z^*}\phi_0
~=~
-\sqrt{\frac{N}{\rho}}~
\left(
\frac{df}{dr}-\frac{f}{r}
\right)
e^{2i\theta}~.
\label{uv0}
\end{eqnarray}
Here $z=x+iy=re^{i\theta}$. This mode is normalized in the usual way 
$\langle u_0|u_0\rangle-\langle v_0|v_0 \rangle=1$. When truncated only 
to the zero mode the Bogoliubov expansion 
(\ref{expansion}) becomes
\begin{equation}
\phi_b \approx
\phi_0 -
{\cal Z} 
\frac{\partial}{\partial z}   \phi_0 - 
{\cal Z}^*   
\frac{\partial}{\partial z^*} \phi_0 ~.
\label{expansion0}
\end{equation}
This expression can be interpreted as the leading terms of a
Taylor expansion in powers of the complex vortex position ${\cal
Z}={\cal X}+i{\cal Y}=\sqrt{\frac{2}{\rho}}~b_0$. To see quantum 
fluctuations of ${\cal Z}$ we must go to the quantum version of the
$N$-conserving Bogoliubov theory.

Any $N$-particle state $|\psi\rangle$ can be written as a
quantum superposition over condensates with different
condensate wave functions $\phi_b$
\begin{eqnarray}
|\psi\rangle~=~
\int d^2b~ \psi(b,b^*)~ 
|N:\phi_b\rangle~.
\label{psi}
\end{eqnarray}
Here $d^2b$ is an abbreviation for $\prod_m d^2b_m$. The state
$|N:\phi_b\rangle$ is a Fock state with $N$ atoms in the wave
function $\phi_b$ from Eq.(\ref{expansion}). For $N\gg 1$ the action 
of the quasiparticle annihilation operator
\begin{equation}
\hat b_m~=~
\frac{ \hat a_0^{\dagger} \hat u_m - 
       \hat a_0 \hat v_m^{\dagger}    }{\sqrt{N}}~,
\end{equation}
with $\hat u_m=\langle u_m|\hat\psi\rangle$ and 
$\hat v_m=\langle v_m^*|\hat\psi\rangle$,
on the state $|\psi\rangle$, turns out to be very simple:
\begin{eqnarray}
&& 
\hat b_m~ |\psi\rangle~= 
\nonumber\\
&&
\int d^2b~ 
\psi(b,b^*)~ 
\hat b_m~
|N:\phi_b\rangle~\stackrel{N\gg1}{\approx}
\nonumber\\
&&
\int d^2b~ 
\left[
\frac{1}{\sqrt{2}}
\left(
b_m+\frac{\partial}{\partial b_m^*}
\right)
\psi(b,b^*)
\right]~ 
|N:\phi_b\rangle~.
\end{eqnarray}
The annihilation (creation) operator $\hat b_m^{(\dagger)}$ acts
on a wave function $\psi(b,b^*)$ as a differential operator. A
`$b$-representation' of the operators is
\begin{eqnarray}
&&
\hat b_m~=~
\frac{1}{\sqrt{2}}
\left(
b_m+\frac{\partial}{\partial b_m^*}
\right)~,
\\
&&
\hat b_m^{\dagger}~=~
\frac{1}{\sqrt{2}}
\left(
b_m^*-\frac{\partial}{\partial b_m}
\right)~.
\end{eqnarray}
It is easy to check, that 
$[\hat b_m,\hat b_n^{\dagger}]=\delta_{mn}$.

The Bogoliubov vacuum state $|0_b\rangle$ is the state without any Bogoliubov 
quasiparticles, so its wave function is annihilated by all the operators 
$\hat b_m$: 
\begin{equation}
\hat b_m|0_b\rangle=0,
\end{equation}
it implies:
\begin{equation}
\psi_0(b,b^*)~\sim~
\prod_m e^{ - b^*_m b_m }~\equiv~
e^{-b^*b}~.
\label{psi_0}
\end{equation} 
If $b_m^{\dagger}$ is applied $l_m$ times to the vacuum state in
order to create $l_m$ quasiparticles, then one gets the
$l_m$-quasiparticle wave function $(\sqrt{2}b_m^*)^{l_m}~e^{-b^*b}$. 
It is quite remarkable that these wave functions look just like
the orbitals of the lowest Landau level. This may be a little bit
surprising because after all, they are the eigenstates of the
quasiparticle harmonic oscillators and not of any particles in a
magnetic field. Also, more generally, a quasiparticle Fock state
with definite numbers of different quasiparticles has a
characteristic `polynomial $\times$ exponential' form
$\left( \prod_m~ (b_m^*)^{l_m} \right)~e^{-b^*b}~$.
A general wave function is $\psi(b,b^*)=P(b^*)e^{-b^*b}$
with an arbitrary polynomial in $b^*$'s. 

As we already could see in Eq.(\ref{expansion0}), in the case of our
planar vortex one of the fictitious particle coordinates $b_0$
materializes as a complex vortex position ${\cal Z}=\sqrt{\frac{2}{\rho}}b_0$. 
The Bogoliubov vacuum wave function (\ref{psi_0}) for this
mode becomes
\begin{equation}
e^{- \frac12 \rho {\cal Z}^*{\cal Z} }~
\label{gauss_b}
\end{equation}
and it is identical with the wave function (\ref{gauss_eff}). In contrast
to Eq.(\ref{gauss_eff}) the wave function (\ref{gauss_b}) is derived from
first principles without any reference to the effective action
(\ref{Seff}).

  In the following we use the same first principles to show that
Eqs.(\ref{gauss_eff}) and (\ref{gauss_b}) are misleading. It does not make sense
to truncate the theory to the vortex zero mode because this mode is not
independent from other Bogoliubov modes.

%%%%%%%%%%%%%%%%%%%%%%%%%%%%%%%%%%%%%%%%%%%%%%%%%%%%%%%%%%%%%%%%%%%%
\section{ A contradiction  }
%%%%%%%%%%%%%%%%%%%%%%%%%%%%%%%%%%%%%%%%%%%%%%%%%%%%%%%%%%%%%%%%%%%%

The quantum fluctuations described by Eq.(\ref{gauss_b}) result
in quantum depletion of atoms from the condensate wave function
$\phi_0(\vec x)$. The density of atoms depleted from the condensate
in the state (\ref{psi}) with the wave function (\ref{gauss_b})
truncated to the vortex mode can be worked out as
\begin{equation}
d\rho_0(\vec x)~
\stackrel{?}{=}
~|u_0(\vec x)|^2+|v_0(\vec x)|^2~.
\label{drho0}
\end{equation}
Here $u_0$ and $v_0$ come in a symmetric way like in Eq.(\ref{expansion}).
This $d\rho_0(\vec x)\neq0$ at $r=0$, the non-zero contribution is
coming from the term $|u_0(\vec x)|^2$. The fluctuating vortex position
results in a non-zero density of depleted atoms inside vortex core. This
is quite natural: a vortex with a fixed position has an empty core but
when the position is uncertain then the average density of atoms inside
the core in non-zero. There is a one to one correspondence between the
magnitude of position fluctuations and the density of depleted atoms in
the core.

On second thought, Eq.(\ref{drho0}) appears strange. After all the
density of depleted atoms in $|0_b\rangle$ state, using the Bogoliubov theory, is given by the sum
$\sum_{m=0}^{\infty}|v_m(\vec x)|^2$ which contains all $v_m$'s but
no $u_m$'s. When truncated only to the vortex mode the sum becomes
\begin{equation}
d\rho_0(\vec x)~
\stackrel{?}{=}
~|v_0(\vec x)|^2~
\label{drho00}
\end{equation}
in apparent contradiction with Eq.(\ref{drho0}). Both Eq.(\ref{drho0})
and Eq.(\ref{drho00}) result from a truncation of the Bogoliubov theory
to the vortex mode. However, the two truncations are done in a different
way. The resulting contradiction shows that such a truncation does not 
make sense. 

The reason why it makes no sense is that different Bogoliubov modes
$(u_m,v_m)$ are not independent from each other. Their wave functions are
not mutually orthogonal in the usual sense: for example, in general
$\langle u_m|u_n\rangle\neq 0$ for $m\neq n$. The wave function of the
vacuum (\ref{psi_0}) factorizes implying independence of different
$b_m$'s, but the non-trivial correlations between different Bogoliubov
modes are hidden in the $N$-particle condensate $|N:\phi_b\rangle$ making
the mode independence rather elusive.
 
%%%%%%%%%%%%%%%%%%%%%%%%%%%%%%%%%%%%%%%%%%%%%%%%%%%%%%%%%%%%%%%%%%%%%%%%
\section{ Phonons drive vortex fluctuations }
%%%%%%%%%%%%%%%%%%%%%%%%%%%%%%%%%%%%%%%%%%%%%%%%%%%%%%%%%%%%%%%%%%%%%%%%

\begin{figure}[ht]
\vspace*{-0.5cm}
\epsfig{file=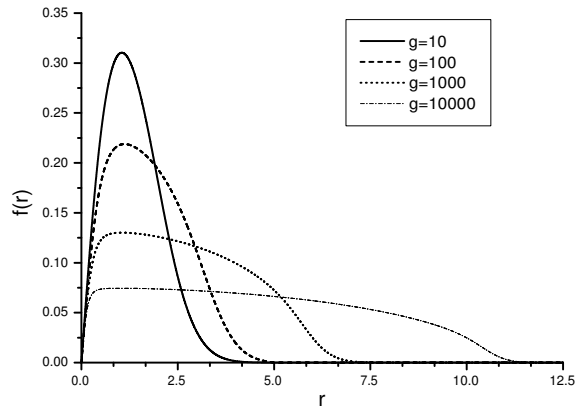, width=8.5cm, clip=}
\vspace*{-0.9cm}
\caption{ Profiles of the vortex wave function for several values of
the interaction strength $g$. The solutions were obtained by relaxation
to the ground state with vortex symmetry. Except for the vortex core
at $r\approx 0$, the profiles $f(r)$ for $g=1000,10000$ match well
with the Thomas-Fermi profile $f_{\rm TF}(r)=\sqrt{\frac{R^2-r^2}{2g}}$.
Here $R=(4g/\pi)^{1/4}$ is the Thomas-Fermi radius in a two dimensional 
trap. 
}
\label{f(r)}
\end{figure}

In contrast to Eq.(\ref{drho0}) equation (\ref{drho00}) shows vanishing 
density of depleted atoms at $r=0$, i.e. there are no vortex position
fluctuations originating from the vortex mode. At the same time the
contradiction between Eqs.(\ref{drho0}) and (\ref{drho00}) shows that
we cannot judge on position fluctuations using any of the two truncations,
but we have to use the full expression
\begin{equation}
d\rho(r=0)~=~\sum_{m=0}^{\infty} |v_m(r=0)|^2
\label{drhor=0}
\end{equation}    
including all Bogoliubov modes. We know that the vortex mode ($m=0$) has
no contribution to $d\rho(r=0)$. To work out the contribution from the
other modes in a controlled way we put the system in a 2D harmonic trap
with a frequency $\omega$. In the dimensionless trap units the stationary
Gross-Pitaevskii equation becomes
\begin{equation}
\mu\phi=
-\frac12\nabla^2\phi+
\frac12 r^2 \phi+
g |\phi|^2\phi~.
\end{equation}
Its vortex solutions $\phi_0=f(r)e^{i\theta}$ for 
several different values of $g$ are shown in trap units in Fig.\ref{f(r)}. 

The Bogoliubov modes are solutions of the Bogoliubov-de Gennes 
equations  
\begin{eqnarray}
\omega_m      u_m &=&
{\cal H}_{GP} u_m +
g\phi_0^2      v_m ~,
\nonumber\\
-\omega_m      v_m &=&
{\cal H}_{GP} v_m +
g(\phi_0^*)^2  u_m ~.
\label{BdG}
\end{eqnarray}
Here ${\cal H}_{GP}=-\frac12\nabla^2+\frac12 r^2+2g|\phi_0|^2-\mu$. 
The symmetry of $\phi_0\sim e^{i\theta}$ implies that
\begin{eqnarray}
&&
u(r,\theta)~=~U(r)~e^{il\theta+2i\theta}~,
\nonumber\\
&&
v(r,\theta)~=~V(r)~e^{il\theta}~.
\end{eqnarray}
The Bogoliubov modes can be classified into different multiplets with 
definite
angular momentum $l$. As only the modes with $l=0$ can contribute to
$d\rho(r=0)$ in Eq.(\ref{drhor=0}), from now on we focus only on the $l=0$
multiplet. 

The Bogoliubov-de Gennes equations were numerically solved for $l=0$ using
the vortex solutions $\phi_0$ (Fig.\ref{f(r)}). In the numerical
calculations the envelopes $U(r)$ and $V(r)$ were expressed as finite
linear combinations of the 2D harmonic oscillator eigenfunctions up to
certain energy cut-off. All the numerical quantities presented below were
checked for convergence with the increasing energy cut-off. We also tested
our numerical code in the limit of $g\ll 1$ where some perturbative
analytical results can be compared with numerics. The Bogoliubov modes of
the $l=0$ multiplet can be classified as follows.

\begin{itemize}

\item The anomalous mode which is a harmonic trap counterpart of the
vortex zero mode (\ref{uv0}). In the trap the mode acquires a non-zero
negative frequency $\omega_a<0$ , where $\omega_a \rightarrow0$ in the limit of ${g\rightarrow\infty}$\cite{Fetter}. It was tested in \cite{rzaz}.

\item The collective oscillations mode with the trap frequency $\omega_1=1$, 

\begin{eqnarray}
&&
u_1~=~
\left(  
\frac{df}{dr}-\frac{f}{r}-rf
\right)
e^{2i\theta}~,
\nonumber\\
&&
v_1~=~
\left(
\frac{df}{dr}+\frac{f}{r}+rf
\right)~.
\end{eqnarray}
The frequency $\omega_1$ of the numerical counterpart of this exact
mode was kept within $0.01$ of the exact $\omega_1=1$. Contribution
of the trap mode to the density of depleted atoms in the core can be
immediately estimated as
\begin{equation}
|v_1(r=0)|^2~=~
\left|\frac{df}{dr}(r=0)\right|^2~\approx~
\frac{1}{R\xi}~=~
{\cal O}(g^0)~.
\end{equation}
This contribution does not depend on $g$. For large $g$ it is negligible
as compared to phonon modes (see next point). Even for a finite $g$ this trap mode is a
collective excitation of the center of mass of the atomic cloud which has
nothing to do with the motion of the vortex with respect to the
condensate. For these two reasons we do not include the trap mode in
$d\rho(r=0)$ even for finite $g$.

\item Phonon modes with positive frequencies $\omega_{m>1}>0$.
These modes give a nonzero contribution to $d\rho(r=0)$ which
grows with $g$ as $\sqrt{g}$. The phonon contribution 
dominates for large $g$ and it results in genuine fluctuation
of the vortex position with respect to the condensate. Rather
surprisingly, fluctuation of vortex position, if any, turns out
to come from phonons and not from the vortex mode.

\end{itemize}

\begin{figure}[ht]
\vspace*{-0.5cm}
\epsfig{file=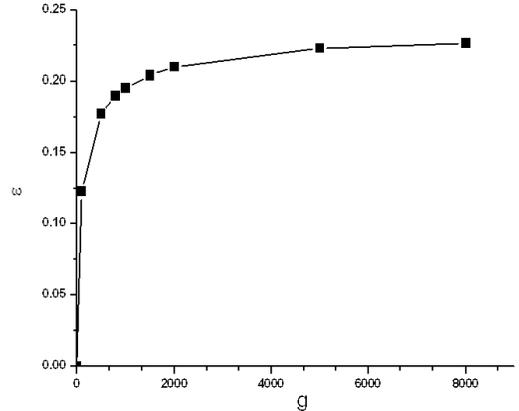, width=8.5cm, clip=}
\vspace*{-0.9cm}
\caption{ $\varepsilon$ is a ratio of the density of depleted
atoms inside the vortex core at $r=0$ to the density of depleted atoms
in a uniform condensate without any vortex (or depletion outside the
vortex core). Small $\varepsilon$
means that the non-condensed atoms are quite efficiently expelled
from the core by the centrifugal force. $\varepsilon$ stabilizes
around $0.23$ for large $g$ when the Thomas-Fermi condensate
can be very well approximated by a uniform condensate, at least
locally close to the vortex in the center of the trap. 
$\varepsilon\approx 0.23$ is the asymptotic value of the reduction 
factor in a uniform condensate. }
\label{varepsilon}
\end{figure}

We compared the dominating phonon contribution 
$d\rho_{\rm phonons}(r=0)=\sum_{m\geq 2}|v_m(r=0)|^2$
to the quantum depletion in a uniform condensate without any vortex.
If the density of the uniform condensate is equal to the density
of our condensate near the center of the trap, then the
quantum depletion in the uniform condensate is 
$d\rho_{\rm uniform}=g^{1/2}/4\pi^{3/2}$. We compare our numerical value
of $d\rho_{\rm phonons}(r=0)$ to this uniform depletion in 
Fig.\ref{varepsilon} where we plot the reduction factor
\begin{equation}
\varepsilon~=~
\frac{d\rho_{\rm phonons}(r=0)}{d\rho_{\rm uniform}}~.
\end{equation}
For large $g$ the reduction factor $\varepsilon$ stabilizes at
approximately $0.23$. We conclude that the density of depleted atoms
inside the vortex core is a factor of $0.23$ less than the same depletion
outside the vortex. Apparently the centrifugal force due to vortex
rotation is quite efficient in throwing away from the core not only the 
condensate but
also the non-condensed atoms.

To summarize we find that for a large $g$ the density of depleted atoms in
the core is dominated by the phonon contribution and is approximately
given by
\begin{equation}
d\rho(r=0)~\approx~
0.23~
\frac{g^{1/2}}{4\pi^{3/2}}~.
\end{equation}
It is time to compare this phonon contribution with the simple estimate
(\ref{drho0}) based on the truncation to the vortex mode. In the
Thomas-Fermi limit we estimate
$|u_0(r=0)|^2=\frac{1}{4\pi}\frac{N}{\rho}\left|\frac{df}{dr}(r=0)\right|^2\approx
0.090 g^{1/2}$. The ratio of the true to the naive density reads
\begin{equation}
\delta~=~
\frac{d\rho(r=0)}{|u_0(r=0)|^2}~\approx~
0.11~.
\end{equation}
We conclude that in the Thomas-Fermi limit the actual fluctuations
of the vortex position are an order of magnitude less than
the simple estimate based on quantization of the effective action
or equivalent truncation of the Bogoliubov theory to the vortex mode.

%%%%%%%%%%%%%%%%%%%%%%%%%%%%%%%%%%%%%%%%%%%%%%%%%%%%%%%%%%%%%%%%%%%%%%%%%%%
\section{ Summary }
%%%%%%%%%%%%%%%%%%%%%%%%%%%%%%%%%%%%%%%%%%%%%%%%%%%%%%%%%%%%%%%%%%%%%%%%%%%

This paper has the following logic. We start from an effective
action for vortex dynamics in two dimensions, quantize it and obtain an
effective wave function for the position of the vortex. Dispersion of this
gaussian wave function seems to estimate quantum fluctuations in vortex
position. Next we introduce the number conserving Bogoliubov theory and
rederive the same wave function. This establishes a connection between the
effective action quantization and the Bogoliubov theory: the former gives
the same result as the latter when the latter is truncated to the vortex
mode. Next, in the framework of the Bogoliubov theory we show that the
truncation is not unique and that vortex position fluctuations critically
depend on the way of truncating to the vortex mode. It turns out that the
vortex mode in the Bogoliubov theory is not independent from the phonon
modes so that all modes have to be included in order to correctly estimate
quantum fluctuations.  Rather surprisingly, in the full untruncated
Bogoliubov theory quantum fluctuations of vortex position come from phonon
modes and not from the vortex mode. What is more these phonon-induced
quantum fluctuations are an order of magnitude less than the simple
estimate based on the truncation to the vortex mode or on the quantization
of the effective action. We want to stress this result means only that the naive quantization of the effective action (\ref{Seff}) does not work properly, not that the effective action is incorrect.

We also found that the density of depleted atoms inside a vortex core is a factor
of $0.23$ less than outside the core.

%%%%%%%%%%%%%%%%%%%%%%%%%%%%%%%%%%%%%%%%%%%%%%%%%%%%%%%%%%%%%%%%%%%%%%%%%%%
\section{ Acknowledgements } 

We are indebted to Krzysztof Sacha for numerous stimulating discussions
and comments. J.D. was supported in part by the KBN grant
PBZ-MIN-008/P03/2003.
%%%%%%%%%%%%%%%%%%%%%%%%%%%%%%%%%%%%%%%%%%%%%%%%%%%%%%%%%%%%%%%%%%%%%%%%%%%

\end{document}